\documentclass[sigconf]{acmart}

\copyrightyear{2026}
\acmYear{2026}
\setcopyright{cc}
\setcctype{by}
\acmConference[SIGIR '26] {Proceedings of the 49th International ACM SIGIR Conference on Research and Development in Information Retrieval}{July 20--24, 2026}{Melbourne, VIC, Australia.}
\acmBooktitle{Proceedings of the 49th International ACM SIGIR Conference on Research and Development in Information Retrieval (SIGIR '26), July 20--24, 2026, Melbourne, VIC, Australia}
\acmISBN{979-8-4007-2599-9/2026/07}
\acmDOI{10.1145/3805712.3809591}
\usepackage{mathtools}
\usepackage{lipsum}
\usepackage{graphicx}
\usepackage{caption, subcaption}
\usepackage{url}
\usepackage{amsmath, amsfonts, amsthm, bm}
\usepackage{soul}
\usepackage{tabularx, multirow}
\usepackage{dsfont}
\usepackage[ruled,vlined]{algorithm2e}
\usepackage{booktabs}
\usepackage{xcolor, colortbl}
\usepackage{xspace}
\usepackage{resizegather}
\usepackage{makecell}
\usepackage[normalem]{ulem}
\usepackage{CJKutf8}
\usepackage{balance}
\usepackage{enumitem}
\usepackage{subcaption}

\newcolumntype{X}{>{\centering\arraybackslash}m{0.05\linewidth}}
\newcolumntype{Y}{>{\raggedleft\arraybackslash}m{0.24\linewidth}}
\newcolumntype{Z}{>{\raggedleft\arraybackslash}m{0.14\linewidth}}
\newcolumntype{Q}{>{\raggedright\arraybackslash}m{1\linewidth}}
\newcolumntype{L}{>{\raggedright\arraybackslash}m{0.56\linewidth}}
\newcolumntype{P}{>{\raggedright\arraybackslash}m{0.98\linewidth}}
\begin{document}

\title{MVIGER: Multi-View Variational Integration of Complementary~Knowledge for Generative Recommender}


\author{Tongyoung Kim}
\affiliation{%
  \institution{Yonsei University}
  \city{Seoul}
  \country{Republic of Korea}}
\email{dykim@yonsei.ac.kr}

\author{Soojin Yoon}
\affiliation{%
 \institution{Yonsei University}
  \city{Seoul}
  \country{Republic of Korea}}
  \email{soojiny@yonsei.ac.kr}

\author{SeongKu Kang}
\affiliation{%
 \institution{Korea University}
  \city{Seoul}
  \country{Republic of Korea}}
    \email{seongkukang@korea.ac.kr}

\author{Jinyoung Yeo}
\affiliation{%
  \institution{Yonsei University}
  \city{Seoul}
  \country{Republic of Korea}}
\email{jinyeo@yonsei.ac.kr}

\author{Dongha Lee}
\authornote{Corresponding author}
\affiliation{%
  \institution{Yonsei University}
  \city{Seoul}
  \country{Republic of Korea}}
\email{donalee@yonsei.ac.kr}

\newcommand{\tsne}{TSNE\xspace}

\newcommand{\ceid}{\textsc{CeID}\xspace}
\newcommand{\seid}{\textsc{SeID}\xspace}
\newcommand{\conf}{\textsc{Conf}\xspace}
\newcommand{\cons}{\textsc{Cons}\xspace}

\newcommand{\rqvae}{RQ-VAE\xspace}

\newcommand{\lightgcn}{LightGCN\xspace}
\newcommand{\grurec}{GRU4Rec\xspace}
\newcommand{\bertrec}{Bert4Rec\xspace}
\newcommand{\sasrec}{SASRec\xspace}
\newcommand{\sthreerec}{S\textsuperscript{3}-Rec\xspace}
\newcommand{\tfive}{T5\xspace}
\newcommand{\pfive}{P5\xspace}
\newcommand{\pfivesid}{P5 (SemID)\xspace}
\newcommand{\pfivecid}{P5 (CID)\xspace}
\newcommand{\pfivecsid}{P5 (CID+SemID)\xspace}
\newcommand{\pfiveciid}{P5 (CID+IID)\xspace}
\newcommand{\tiger}{TIGER\xspace}
\newcommand{\eager}{EAGER\xspace}
\newcommand{\pfiveseid}{P5 (\seid)\xspace}
\newcommand{\pfiveceid}{P5 (\ceid)\xspace}
\newcommand{\pfivecseid}{P5 (\ceid + \seid)\xspace}
\newcommand{\proposed}{\textsc{MVIGER}\xspace}
\newcommand{\singleproposed}{\textsc{MVIGER} (Single)\xspace}
\newcommand{\multiproposed}{\textsc{MVIGER} (Multi)\xspace}

\newcommand{\idnotok}[2]{\text{ID}_{#1,#2}\xspace}
\newcommand{\idtok}[2]{\langle\text{ID}_{#1,#2}\rangle\xspace}
\newcommand{\ceidtok}[2]{\langle\text{C\textsc{e}ID}_{#1,#2}\rangle\xspace}
\newcommand{\seidtok}[2]{\langle\text{S\textsc{e}ID}_{#1,#2}\rangle\xspace}

\newcommand{\veryshortarrow}[1][3pt]{\mathrel{%
\hbox{\rule[\dimexpr\fontdimen22\textfont2-.2pt\relax]{#1}{.4pt}}%
\mkern-4mu\hbox{\usefont{U}{lasy}{m}{n}\symbol{41}}}}
\makeatletter
\definecolor{Gray}{gray}{0.9}

\newcommand{\amazon}{Amazon\xspace}
\newcommand{\amazonb}{Amazon Beauty\xspace}
\newcommand{\amazons}{Amazon Sports\xspace}
\newcommand{\amazont}{Amazon Toys\xspace}
\newcommand{\yelp}{Yelp\xspace}

\newcommand{\reduce}[1]{\textls[-50]{#1}}
\newcommand{\smallsection}[1]{{\vspace{0.05in} \noindent \bf {#1.\hspace{5pt}}}}
\newcommand{\proposedcsid}{\textsc{PC-Rec} (CID, SemID)\xspace}
\newcommand{\proposedcseid}{\textsc{PC-Rec} (\textsc{CeID}, \textsc{SeID})\xspace}

\begin{abstract}
Language Models (LMs) have been widely used in recommender systems to incorporate textual information of items into item IDs, leveraging their advanced language understanding and generation capabilities.
Recently, generative recommender systems have utilized the reasoning abilities of LMs to directly generate index tokens for potential items of interest based on the user's interaction history.
To inject diverse item knowledge into LMs, prompt templates with detailed task descriptions and various indexing techniques derived from diverse item information have been explored.
This paper focuses on the inconsistency in outputs generated by variations in input prompt templates and item index types, even with the same user’s interaction history.
Our in-depth quantitative analysis reveals that preference knowledge learned from diverse prompt templates and heterogeneous indices differs significantly, indicating a high potential for complementarity.
To fully exploit this complementarity and provide consistent performance under varying prompts and item indices, we propose \proposed, a unified variational framework that models selection among these information sources as a categorical latent variable with a learnable prior.
During inference, this prior enables the model to adaptively select the most relevant source or aggregate predictions across multiple sources, thereby ensuring high-quality recommendation across diverse template-index combinations.
We validate the effectiveness of \proposed on three real-world datasets, demonstrating its superior performance over existing generative recommender baselines through the effective integration of complementary knowledge.

\end{abstract}



\maketitle

\section{Introduction}
\label{sec:intro}
In the era of information overload, recommender systems have become essential tools to help users efficiently discover items aligned with their preferences and interests.
Sequential recommendation~\cite{kang2018self, sun2019bert4rec, hidasi2018recurrent, tang2018personalized, zhou2020s3} specifically focuses on predicting user's subsequent interactions by leveraging their historical interaction sequences, aiming to deliver more personalized suggestions.
With the remarkable advancement of LMs, recent progress in sequential recommender systems has been driven by utilizing LM, particularly by leveraging their strong text understanding, reasoning, and generation capabilities~\cite{dai2023uncovering, bao2023tallrec, yuan2023go, li2023exploring, zhang2023collm, hou2024large}.
To effectively formulate the recommendation task within generative LM, existing methods have adopted the generative retrieval approach~\cite{rajput2024recommender, hua2023index, zheng2024adapting, zhu2024cost, wang2024eager, si2024generative}, which decodes the item identifiers (e.g., titles, attributes, descriptions and numeric IDs) using historical interaction data of the user as input prompt.
Based on this architecture, several foundational models have been proposed, such as \pfive~\cite{geng2022recommendation} and M6Rec~\cite{cui2022m6}, which are specifically fine-tuned for diverse recommendation tasks.

\begin{figure}[t] 
    \centering 
    \includegraphics[width=\linewidth]{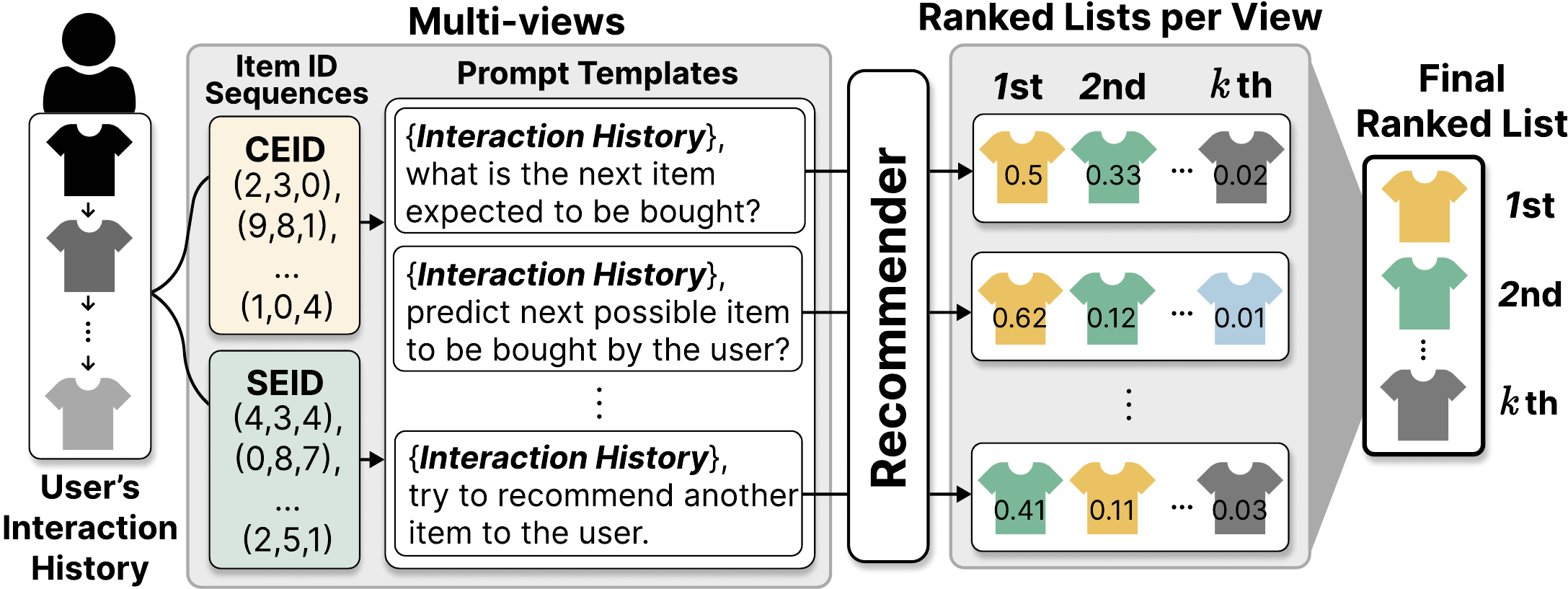} 
    \caption{Inconsistent predictions for the same user's history due to variations in prompt templates and item index, motivating the need for integrating complementary knowledge.}
    \label{fig:intro} 
\end{figure}

To effectively inject item knowledge into LM-based recommenders via item identifier tokens, several studies have proposed hierarchical item indices~\cite{rajput2024recommender, hua2023index, zhu2024cost, wang2024eager, si2024generative}, in which the hierarchical structure encodes diverse aspects of item knowledge.
Most existing studies have tried to encode \textit{semantic} information of items, including textual descriptions~\cite{rajput2024recommender, lin2023multi, zhu2024cost, si2024generative} or predefined category~\cite{hua2023index, lin2023multi} provided by item metadata, into the hierarchical index structure.
Additionally, inspired by the concept that items with more frequent co-occurrence are more similar, several indexing methods have been proposed to distill \textit{collaborative} information into the index structure;
for example, previous works employ clustering on the item co-occurrence matrix \cite{hua2023index} by using matrix factorization or pretrained collaborative filtering model's item embedding~\cite{wang2024eager}.

In this work, we focus on varied outputs generated by a recommender system, depending on (1) input prompt templates that describe the detailed instruction for the sequential recommendation task, and (2) types of hierarchical item indices that are used to represent each item as multiple tokens.
The input prompts consist of templates and item indices, and variations in these lead to inconsistent top-$k$ predictions from the model, even when using the same user's interaction history. 
This sensitivity implies that the same model might generate different outputs simply by changing how the task is described or how items are indexed, making it challenging to provide consistent and stable recommendations. 
For example, Figure~\ref{fig:intro} shows inconsistent predictions of the same interaction history when different prompts are used.

Interestingly, our preliminary analysis shows that the model's diverse outputs capture complementary knowledge, indicating that different prompt templates and index types leverage diverse sources of information. 
Specifically, less than 50\% of the correctly predicted items are shared among the retrieval results obtained with different item index types.
Furthermore, the model captures diverse contexts from various prompt templates, leading to varied results that encompass complementary predictions by over 5\%.
Motivated by these findings, we treat the diverse combinations of prompt templates and index types as a latent categorical variable, offering a single conceptual framework to capture their complementary nature.

Based on this motivation, we propose \proposed, a unified variational integration method designed to integrate diverse template-index combinations as distinct views within a single language model. 
By introducing a categorical latent variable to select among diverse views and modeling its prior and posterior distributions, \proposed systematically learns how to integrate complementary information from different prompt templates and heterogeneous index types.
Through this probabilistic integration, the model is able to fully leverage complementary knowledge encoded in multiple views, capturing a broader range of user preferences and item relationships than any single view.
By learning a prior distribution over possible template-index choices conditioned on each user's history, \proposed can flexibly control how to utilize diverse information from the views during inference.
Specifically, the model can select the most probable view or aggregate the predictions across multiple views, offering flexibility to balance speed and performance.

Our empirical evaluations in three real-world datasets clearly demonstrate that \proposed significantly outperforms state-of-the-art methods, generating more consistent recommendations by effectively integrating complementary knowledge from diverse prompts and indices.
We also provide extensive experiments, including quantitative evaluation, ablation studies, and exploratory analysis, to verify the effectiveness of \proposed in sequential recommendation.

Our contributions can be summarized as follows:
\begin{itemize}
[leftmargin=1.2em, itemsep=2pt, topsep=2pt]
    \item \textbf{Novel problem formulation.} We identify the inconsistency in LM-based sequential recommendation caused by variations in prompt templates and item index types, and empirically show that their diverse outputs encode complementary knowledge.
    
    \item \textbf{Unified integration framework.} We introduce \proposed, a unified variational integration method that treats diverse template-index combinations as distinct views, each representing a different prompting context and item-knowledge configuration for the same user, and models their selection with a categorical latent variable.
    We also summarize the multi-view training and inference procedure in Algorithm~\ref{alg:mviger}.
    The algorithm illustrates how multiple user-conditioned views are constructed and how the likelihood and prior networks are optimized.
    
    During training, the variational posterior over views is computed in closed form and treated as an optimal posterior, enabling decoupled updates of the likelihood and prior networks in a variational EM-style procedure.
    At inference, the model supports two strategies: selecting the most probable view for efficient prediction, or aggregating predictions across views using the learned prior to effectively leverage complementary knowledge.
    This framework systematically integrates complementary knowledge within a single probabilistic model.

    \item \textbf{Flexible inference mechanism.} \proposed learns a prior distribution over possible template–index choices conditioned on user history, enabling flexible inference that can select the most probable view or aggregate complementary predictions across multi-view combinations, thereby achieving adaptive trade-offs between efficiency and accuracy at inference time.

\end{itemize}

\section{Related Work}
\label{sec:relatedwork}
In this section, we briefly review the literature on (1) generative retrieval approaches and (2) sequential recommendation methods.

\subsection{Generative Retrieval}
\label{subsec:genret}
Generative retrieval is a new approach where models directly generate relevant content in the database.
This approach leverages the generation capabilities of generative models to enhance information retrieval systems with nuanced and contextually aware retrieval mechanisms.
GENRE~\cite{de2020autoregressive} is introduced for entity retrieval task, which retrieves entities by generating their unique names token-by-token in an autoregressive manner.
This approach mitigates the limitations of dense entity representations, which require large memory footprints and negative data sub-sampling.
DSI~\cite{tay2022transformer} is introduced for document retrieval task, which retrieves document identifiers that are relevant to queries.
It first assigns a structured semantic document ID to each document. 
Then given a query, a unified text-to-text model is trained to autoregressively return the identifiers of the document token-by-token.
The semantic IDs are generated by hierarchical clustering on the document representations.
Subsequently, NCI~\cite{wang2022neural} proposes a new decoder architecture that can take into account position information for DSI~\cite{tay2022transformer}.

\subsection{Sequential Recommendation}
Sequential recommendation aims to predict the next item users will interact with based on the sequence of their previous interactions.

\subsubsection{Traditional Recommendation} 
Early approaches often rely on Markov Chains techniques to model user behavior based on historical interactions \cite{rendle2010factorization}. 
\grurec~\cite{hidasi2015session} firstly uses the Gated Recurrent Unit (GRU) based RNNs for sequential recommendations.
Subsequent methods like SASRec~\cite{kang2018self} employ the self-attention mechanism, akin to decoder-only transformer models, to capture long-range dependencies in sequential data.
Also, transformer-based models, such as BERT4Rec~\cite{sun2019bert4rec} and Transformers4Rec~\cite{de2021transformers4rec}, leverage masking strategies for training.
To further enhance the recommendation performance, recent studies have explored pre-training techniques using self-supervised learning tasks. 
S$^3$-Rec \cite{zhou2020s3} uses pre-training on four self-supervised tasks to improve the quality of item embeddings.
VQ-Rec \cite{hou2023learning} proposes a new approach for transferable sequential recommenders, addressing issues related to the tight binding between item text and representation through contrastive pre-training and cross-domain fine-tuning methods.

Parallel to these developments, variational autoencoder (VAE) frameworks such as cVAE~\cite{li2017collaborative}, Mult-VAE~\cite{liang2018variational} and RecVAE~\cite{shenbin2020recvae} model user-item interactions via latent variables and optimize the evidence lower bound (ELBO).
While these methods enable probabilistic modeling of user-item interactions, they represent user or item preference using a single latent vector and do not consider integrating multiple views of the same user context.
In contrast, our approach introduces multiple latent views for each template-index combination and 
generalizes variational inference to systematically integrate these multiple views within a unified probabilistic model, effectively capturing complementary knowledge.

\subsubsection{Generative Sequential Recommender}
Recently, the generative retrieval approach has been actively employed for sequential recommendation \cite{rajput2024recommender, hua2023index, zheng2024adapting, zhu2024cost, si2024generative, wang2024eager}.
This approach leverages the reasoning and generation capabilities of LMs to directly generate sequences of indices for the target item.
\pfive~\cite{geng2022recommendation} integrates various recommendation tasks into a natural language generation framework using a sequence-to-sequence model.
Using multiple personalized prompt templates, it converts all recommendation data, such as user-item interactions, user descriptions, item metadata, and reviews, into natural language sequences.
\pfive-IDs~\cite{hua2023index} introduces various indexing methods like Collaborative ID, Semantic ID, and Sequential ID, to be used with the P5 model. 
By analyzing and comparing various indexing methods, it also suggests that a hybrid of indexing methods can further enhance performance.
\tiger~\cite{rajput2024recommender} introduces a novel approach for generating item IDs leveraging the textual descriptions of the items.
It utilizes a tree-structured vector quantization (VQ) \cite{van2017neural}, creating a sequence of quantized codewords as item IDs.
TransRec~\cite{lin2023multi} introduces multi-facet identifiers that cover various aspects (e.g., title, attribute) from the textual metadata of items, to enhance the semantic richness of IDs.
LC-Rec~\cite{zheng2024adapting} proposes various semantic alignment tasks (e.g.,  sequential item prediction, item ID-text alignment) to facilitate the integration of item indices into LLMs.
Recent research~\cite{chen2024enhancing, zhu2024cost, si2024generative, zheng2025universal} has further advanced tokenization methods for generative recommendation.

Despite their effectiveness, previous methods have predominantly focused on item indices based solely on textual semantics or collaborative information. 
It is worth noting that few attempts utilize diverse information for indexing.
LETTER~\cite{wang2024learnable} proposes a contrastive learning-based tokenization approach that aligns semantic and collaborative information into a unified item token.
However, it does not aim to leverage these heterogeneous information sources as complementary views for adaptive integration during recommendation.
\eager~\cite{wang2024eager} proposes a two-stream generation architecture leveraging a shared encoder and two separate decoders to decode behavior tokens and semantic tokens.
While \eager demonstrates the benefit of integrating collaborative and semantic information, yet its design is restricted to these two sources.
Extending integration to broader prompt–index combinations as heterogeneous views remains an open direction.

\section{PRELIMINARIES}
\label{sec:preliminary}
In this section, we outline the sequential recommendation task and analyze the extent to which variations in prompt templates and indexing types capture different aspects of knowledge.
This motivates the need for their systematic and probabilistic integration.

\subsection{Problem Formulation}
Given a user-item interaction dataset, let $\mathcal{U}$ and $\mathcal{I}$ denote the set of users and items, respectively.
User-item interactions (e.g., review, click, and purchase) are represented by $H(u)=[i_u^1, i_u^2, \ldots , i_u^L]$, where $u \in \mathcal{U}, i_u^l \in \mathcal{I}$, and $L = |H(u)|$.
The goal of the sequential recommendation task is to predict the next item $i_u^{L+1}$ that a user will be interested in based on the user's interaction history $H(u)$.

In this work, we focus on a LM-based sequential recommendation~\cite{geng2022recommendation}, which takes a user's interaction history in the form of an natural language instruction prompt.
In this setting, $\mathcal{T}$ denotes the set of diverse templates used for prompting the language model, and $P_t$ is the set of prediction (i.e., test interaction) presented in the output top-$k$ ranked list when using template $t$.

\subsection{Complementarity of Templates and Indices}
\label{subsec:motivanal}
We conduct an in-depth analysis of how much complementary knowledge is encoded in the generated ranked lists when predicting the next item using different (1) prompts and (2) index types.

\subsubsection{Experimental settings for analysis}
To explore the complementarity of ranking results in sequential recommendation models, we finetune the \pfive model~\cite{geng2022recommendation} to predict the next item using diverse prompt templates and heterogeneous types of item index.
The model captures context from the input prompt, which is constructed from an interaction sequence of item indices based on a prompt template, and then generates a ranked list for the next item.
Since the input context serves as preference knowledge for generating a ranked list, different contexts produce diverse results.

To consider diverse contexts from different prompts, we adopt the 10 prompt templates proposed in the original \pfive~\cite{geng2022recommendation}, each introducing variations in vocabulary or sequence ordering, thus providing distinct contextual signals to the model.
For item representation, we construct two types of indices for each item: a collaborative index (\ceid), derived from user-item interaction embeddings, and a semantic index (\seid), obtained from item metadata such as titles and descriptions.
Both types of embeddings are discretized into hierarchical index tokens via residual quantized variational autoencoder \rqvae~\cite{lee2022autoregressive}, following~\cite{rajput2024recommender}.
This approach enables us to systematically evaluate the effect of context diversity on the model's ability to capture complementary preference knowledge.
Further details on index generation are provided in Method Section~\ref{subsec:indexing}.

\begin{figure}[h] 
    \centering 
    \includegraphics[width=\linewidth]{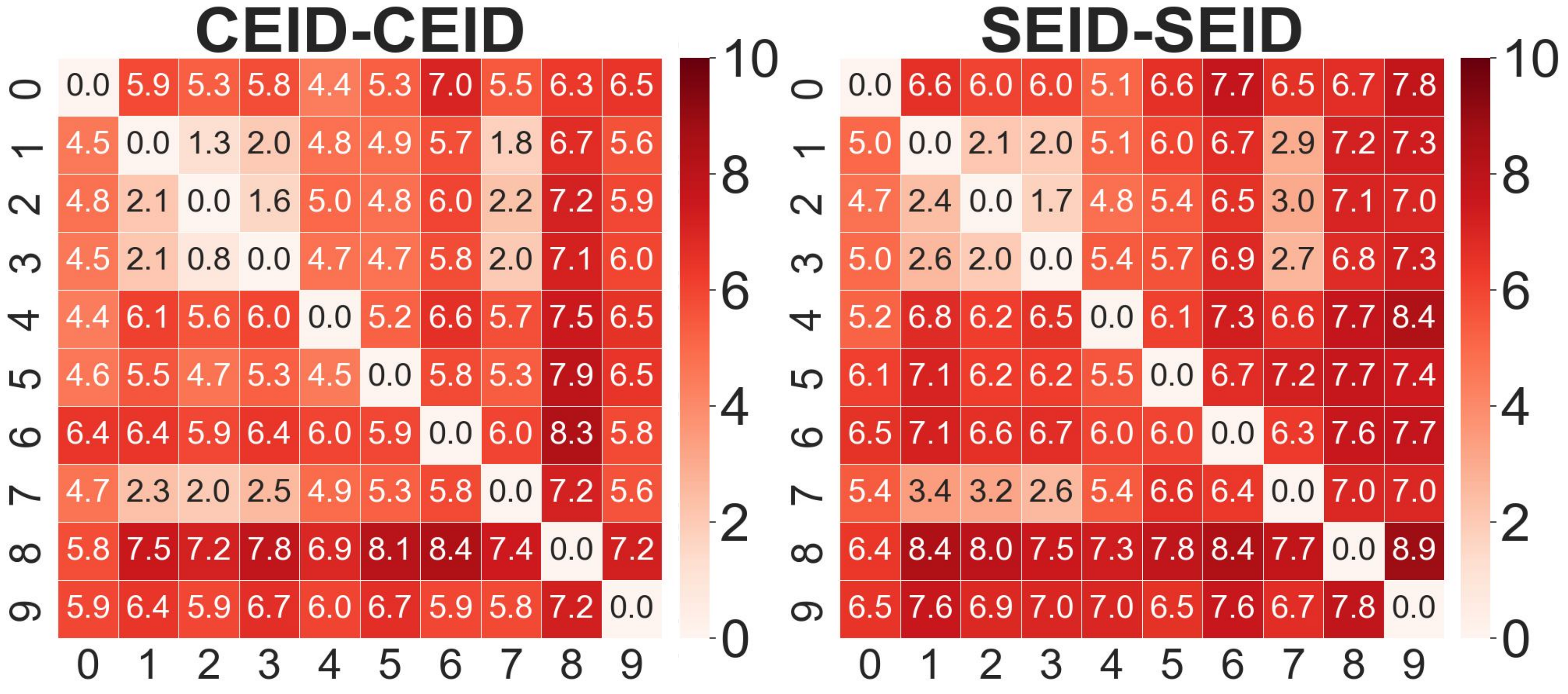} 
    \caption{PER (\%) of \ceid-\ceid and \seid-\seid results from 10 different prompt templates in \amazons.} 
    \label{fig:PER} 
\end{figure}

\subsubsection{Complementarity across prompts}
\label{subsubsec:per}

To investigate the complementarity of results across 10 different prompt templates, we measure Pairwise Exclusive-hit Ratio (PER)~\cite{kang2022consensus} which represents the proportion of correct results exclusively predicted by prompts.
\begin{equation}
    \text{PER}(t_1; t_2) = \frac{|P_{t_1} - P_{t_2}|}{|P_{t_1}|}.
\end{equation}
$\text{PER}(t_1;t_2)$ quantifies the knowledge of user-item relationship correctly captured by template $t_1$ but not by template $t_2$ based on Hit@10 predictions.
We compute the PER values for all pairs among the 10 different templates, and visualize the results in the PER map. 
In Figure~\ref{fig:PER}, there exists complementarity in hit ratio across prompt variations within the same index type (i.e., \ceid and \seid), with minor prompt changes causing up to 5\% performance differences.

\begin{table}[h]
    \centering
    \caption{Average CHR (\%) of \ceid-\seid and \seid-\ceid results from 10 different prompt templates.}
    \label{tab:CHR}
    \resizebox{\linewidth}{!}{
    \begin{tabular}{rcc}
    \toprule
    \textbf{Dataset} & 
    \footnotesize\textbf{CHR\textsubscript{avg}(}$ \bm{\mathcal{T}}_\textbf{\ceid};\bm{\mathcal{T}}_\textbf{\seid}$\textbf{)} &
    \footnotesize \textbf{CHR\textsubscript{avg}(}$
    \bm{\mathcal{T}}_\textbf{\seid}; \bm{\mathcal{T}}_\textbf{\ceid}$\textbf{)}  \\
    \midrule
    {\amazonb} & 48.70 & 50.81 \\ 
    {\amazons}  & 55.64 & 56.81 \\ 
    {\yelp} & 66.90 & 76.93 \\ 
    \bottomrule
    \end{tabular}
    }
\end{table}

\subsubsection{Complementarity across index types}
\label{subsubsec:chr}

We also measure Complementary Hit Ratio (CHR)~\cite{kang2022consensus} to examine the complementarity between the two different index types.
\begin{equation}
    \text{CHR}_\text{avg}(\mathcal{T}_1; \mathcal{T}_2) = \frac{1}{|\mathcal{T}_1|} \sum_{t\in \mathcal{T}_1} \frac{|\bigcup_{t'\in \mathcal{T}_2}P_{t'} - P_t|}{|\bigcup_{t'\in \mathcal{T}_2} P_{t'}|}.
\end{equation}

$\text{CHR}_\text{avg}(\mathcal{T}_1; \mathcal{T}_2)$ quantifies the average of the complementary knowledge that cannot be captured by each of templates in $\mathcal{T}_1$ but can be covered by the predictions from the set of templates in $\mathcal{T}_2$ based on Hit@10 predictions.
A high $\text{CHR}_\text{avg}$ value indicates that prediction results can be greatly complemented by those of the counterpart.
Notably, in Table~\ref{tab:CHR}, we observe that the $\text{CHR}_\text{avg}$ values between the different item index types exceed 48\%. 
This result indicates that the inherent information encoded in the item index allows the recommendation model to acquire different knowledge. 

\subsubsection{Observations}
\label{subsubsec:observation}
Our preliminary experiments demonstrate that the model captures highly diverse contextualized representations (i.e., preference knowledge), resulting in varied inference outcomes due to differences in prompts and item index types.
Firstly, varied prompts introduce nuanced differences in the contextual information considered by the model, leading to diverse outputs.
Secondly, the model acquires significantly different contextualized representations from heterogeneous indices.
This shows that leveraging heterogeneous indices derived from various sources introduces different inductive biases, which enable the model to learn diverse patterns or representations, potentially offering complementary information.
This diversity allows the model to provide more comprehensive recommendations by leveraging multiple perspectives. 

In summary, these findings indicate that systematically integrating complementary information from diverse prompts and item indices has the potential to significantly improve recommendation performance, motivating our proposed integration framework.

\section{Method}
\label{sec:method}
\begin{figure*}[thbp]
    \centering 
    \includegraphics[width=\linewidth]{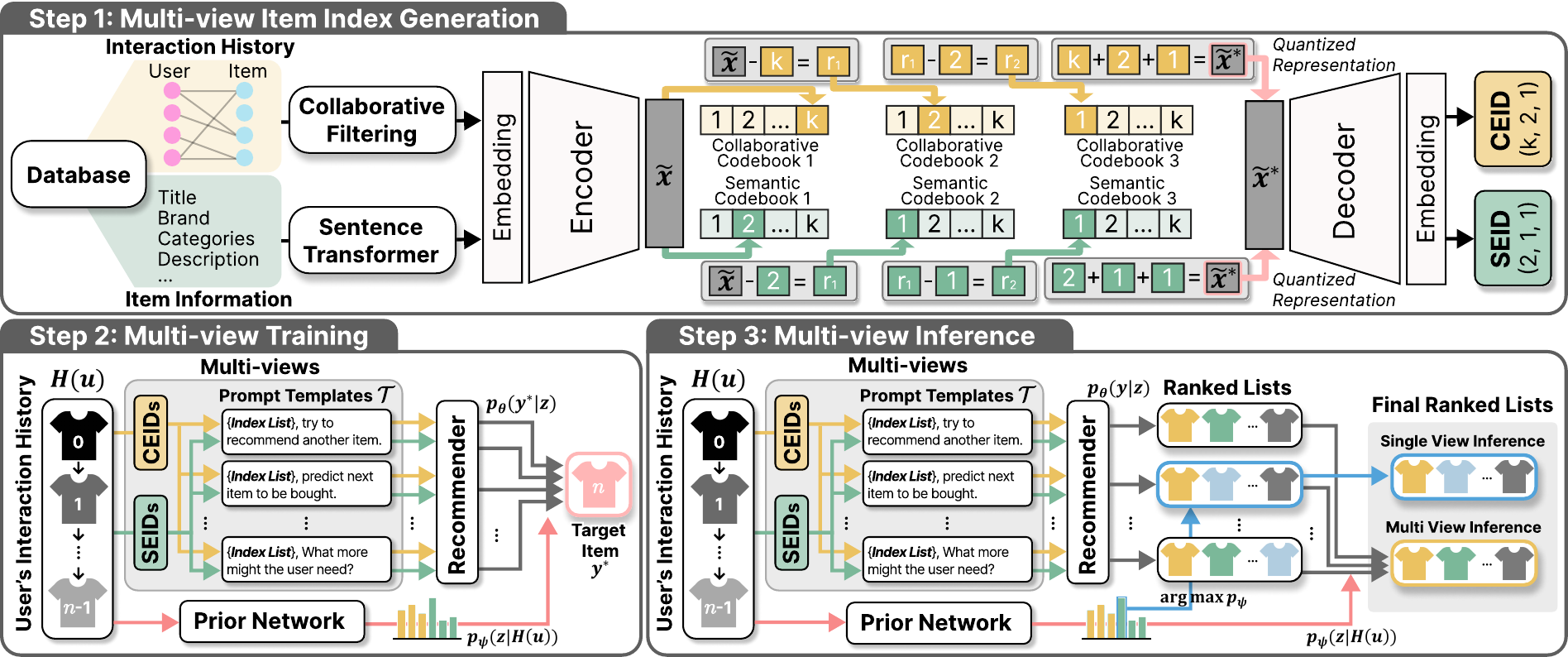} 
    \caption{Overview of the \proposed framework.
    For each user, heterogeneous item indices (e.g., \ceid and \seid) are first constructed based on the user’s interaction history, resulting in multiple latent views with templates.
    \proposed then jointly trains the sequential recommender with a variational prior distribution over these latent views, capturing their relative importance for each user and guiding the integration of information from diverse prompt-template and index combinations.
    At inference, the learned prior adaptively selects or aggregates information from the candidate views, integrating complementary knowledge to generate a final ranked list that is robust and consistent across different template-index settings.}
    \label{fig:framework} 
\end{figure*}

In this section, we describe our proposed \textbf{M}ulti-view \textbf{V}ariational \textbf{I}ntegration of \textbf{GE}nerative \textbf{R}ecommender, \proposed, which systematically integrates complementary knowledge from diverse prompt templates and heterogeneous item indices.
The framework consists of three stages: (1) {heterogeneous} item index generation, {(2) multi-view recommender training, and (3) flexible inference via learned prior.}
The overall framework of \proposed is illustrated in Figure~\ref{fig:framework}.

\subsection{Heterogeneous Item Index Generation}\label{subsec:indexing}
We first construct two distinct item indices for each item, each capturing different aspects of item information derived from collaborative embeddings and semantic embeddings.

\subsubsection{Item embedding generation}
Item embeddings are computed using either \textit{collaborative} or \textit{semantic} information. 
To obtain collaborative embedding of items, we first define a user-item interaction matrix, then optimize a graph convolution-based collaborative filtering (CF) model, such as \lightgcn\footnote{{https://github.com/gusye1234/LightGCN-PyTorch}}~\cite{he2020lightgcn}.
We use this encoder to convert a user's item interaction history into a collaborative embedding. 
Note that \lightgcn is optimized solely on the training data same as that used for sequential recommendation tasks to prevent potential data leakage.
Next, we generate representation sentences for each item by linearizing its metadata (title, brand, categories, description, and attributes).
We then encode these sentences using Sentence-T5 \cite{ni2021sentence} to obtain semantic item embeddings.
To fairly utilize both types of item embedding, we set their dimension sizes to 768, keeping all other hyperparameters as defaults.

\subsubsection{Hierarchical item indexing via residual quantization}
To construct item indices from these embeddings, we apply a residual quantized variational autoencoder (\rqvae)~\cite{lee2022autoregressive}.
\rqvae hierarchically quantizes embeddings into discrete latent representations (codewords), enabling the autoregressive generation of item indices.
Formally, given an item embedding $\mathbf{e}\in \mathbb{R}^d$, the encoder computes a latent representation $\tilde x\in \mathbb{R}^{d'}$.
\rqvae then recursively quantizes latent representation $\tilde x$ at each level $k$:
\begin{equation}
\begin{split}
    c_k &= \operatorname{argmin}_w\Vert r_{k-1}-\mathbf{e}_w^{(k)}\Vert_2^2 \\
    r_{k} &= r_{k-1} - \mathbf{e}_{c_k}^{(k)}, r_0=\tilde x
\end{split}
\end{equation}
At the end of the final level $K$, we obtain the hierarchical item index $(c_1, c_2, ... c_K)$.
The decoder reconstructs the embedding $e^*$ from the quantized representation $\tilde x^*$:
\begin{equation}
    \tilde x^*=\sum_{k=1}^K \mathbf{e}_{c_k}^{(k)}, \mathbf{e}^* = Decoder(\tilde x^*)
\end{equation}
The model is trained with the objective $\mathcal{L}_{\text{RQ-VAE}}$, designed to jointly minimize reconstruction and quantization errors:
\begin{equation}
    \left\lVert \mathbf{e} - \mathbf{e}^*\right\rVert_2^2 + \sum_{k=1}^K \left\lVert \text{sg}\left[r_{k-1}\right] - \mathbf{e}_{c_k}^{(k)}\right\rVert_2^2 + \beta \left\lVert r_{k-1} - \text{sg}[\mathbf{e}_{c_k}^{(k)}]\right\rVert_2^2,
\end{equation}
where the sg[$\cdot$] represents the stop-gradient operator~\cite{van2017neural}, and $\beta$ is a coefficient which is set to 0.25, as done in \cite{lee2022autoregressive}.
We utilize distinct codebooks at each level to attain more refined representations of indices in a coarse-to-fine manner.
The coarse-to-fine indexing enhances autoregressive generation by providing hierarchical representations of items, allowing the model to capture dependencies at multiple levels of abstraction. 
This hierarchical structure enriches the contextual signals available to the model, facilitating more accurate modeling of user preference.

\subsubsection{Handling collisions of item IDs}
Depending on the distribution of item embeddings, codeword collisions may occur, which results in multiple items being mapped to the same ID.
To mitigate these collisions, we introduce an additional token at the final level of the codeword to ensure their uniqueness.
In the event of a collision, we assign a unique identifier (starting from 1) to the last codeword of the items that collide.
Additionally, to maintain consistent codeword lengths, we append the codeword zero (0) to the unique ID of the items that do not collide.

\subsection{Multi-view Recommender Training}
To systematically integrate complementary knowledge from diverse template–index combinations, we define a user-conditioned view set $\mathcal{Z}(u)$, where each view corresponds to a distinct context instantiated by a prompt template and an item index type derived from the interaction history $H(u)$. 
Here, prompt templates provide different contextual formulations, while item index types reflect different knowledge sources, such as collaborative or semantic information. We then introduce a categorical latent variable $z\in\mathcal{Z}(u)$ to model the selection of a view.
Given a user $u$ with interaction history $H(u)$ and a target item $y^*$, \proposed formulates the problem of maximizing the marginal likelihood over latent views:

\begin{equation}
\label{eq:marginal}
    p(y^*|H(u)) = \Sigma_z p_{\theta}(y^*|z) p_\psi(z|H(u)).
\end{equation}

To enable flexible modeling of latent views, we parameterize the prior distribution $p_\psi(z|H(u))$ as a learnable function conditioned on the user's interaction history. 
This allows the model to learn adaptive probabilities over latent views by modeling the distribution of user contexts.
We introduce $q(z \mid H(u), y^*)$, a variational posterior, as a tractable approximation and optimize the marginal likelihood using the Evidence Lower Bound (ELBO), denoted $\mathcal{L}_{\text{ELBO}}$:

\begin{equation}
\label{eq:elbo}
    \mathbb{E}_{q(z|H(u),y^*)}[\log p_{\theta}(y^*|z)] -\text{KL}(q(z|H(u),y^*) || p_{\psi}(z|H(u))).
\end{equation}
To derive the optimal form of the variational posterior, we maximize the ELBO with the constraint that $q(z|H(u),y^*)$ is a valid probability distribution (i.e., it sums up to 1 over $z$).
This can be formulated using the following Lagrangian $\mathcal{F}(q,\lambda)$:
\begin{equation}
    \Sigma_z q(z|H(u),y^*) \log \left[\cfrac{p_{\theta}(y^*|z)  p_{\psi}(z|H(u))}{q(z|H(u),y^*)}\right] + \lambda (1-\Sigma_z q(z|H(u),y^*))
\end{equation}
Taking the derivative of $\mathcal{F}(q,\lambda)$ with respect to variational posterior and setting it to zero gives the optimality condition:
\begin{equation}
\small
    \cfrac{\partial\mathcal{F}}{\partial q(z|H(u),y^*)} = \log \left[\cfrac{p_{\theta}(y^*|z) p_{\psi}(z|H(u))} {q(z|H(u),y^*)}\right] -1 -\lambda = 0.
\end{equation}
Solving this equation for $q(z|H(u),y^*)$, we obtain the closed-form solution for the optimal variational posterior:
\begin{equation}
\label{eq:opt_post}
    q^*(z|H(u),y^*) = \cfrac{p_\theta(y^*|z) p_\psi(z|H(u))}{\Sigma_{z'}p_\theta(y^*|z') p_\psi(z'|H(u))}.
\end{equation}

This result shows that the optimal posterior is proportional to the product of the likelihood and the prior, normalized over all possible views.
Importantly, since the optimal posterior $q^*(z|H(u),y^*)$ is a closed-form solution, it is not parameterized and can be computed directly from the likelihood and the prior distributions.
As a result, when optimizing the model parameters $\theta$ and $\psi$, we treat $q^*$ as a fixed distribution and do not propagate gradients through it.
Under this formulation, the likelihood network is optimized to maximize the expected log-likelihood under $q^*$, while the prior network is optimized to align with the optimal posterior by minimizing the KL divergence $\text{KL}(q^* || p_{\psi}(z|H(u))).$
The training objective is given by:
\begin{equation}
    \mathcal{L}_{ELBO} = \mathbb{E}_{q^*}[\log p_{\theta}(y^*|z)] -\text{KL}(q^* || p_{\psi}(z|H(u))).
\end{equation}

We summarize the overall multi-view training procedure of \proposed in Algorithm~\ref{alg:mviger}.
During training, the optimal posterior over views is computed in closed form and treated as fixed with respect to gradient propagation, allowing the likelihood and prior networks to be optimized according to their respective objectives within the variational framework. 
The likelihood network is implemented using a pretrained \tfive~\cite{raffel2020exploring} model following~\cite{geng2022recommendation}, where the user's interaction history along with each latent view $z$ are provided as input.
For the prior network, the user’s raw interaction history is first encoded into a contextual representation by the T5~\cite{raffel2020exploring} encoder, and this representation is then mapped to a prior distribution over $z$ using a single linear layer.

\begin{algorithm}[h]
\small
\caption{Multi-view Training Procedure}
\label{alg:mviger}
\KwIn{
Training data $\mathcal{D}$; templates $\mathcal{T}$; index types $\mathcal{M}$;\\
likelihood network $\theta$; prior network $\psi$
}
\KwOut{Trained networks $\theta, \psi$ and ranked item list}

\textbf{Notation:} $\mathrm{sg}[\cdot]$ denotes the stop-gradient operator.\;

\textbf{Training Phase}

Initialize likelihood network $\theta$ and prior network $\psi$\;

\While{not converged}{
    Sample a mini-batch $(u, H(u), y^*) \sim \mathcal{D}$\;

    Construct user-conditioned view set:
    \[
    \mathcal{Z}(u) = \{ z=(t,m,H(u)) \mid t \in \mathcal{T},\, m \in \mathcal{M} \}
    \]

    \ForEach{$z \in \mathcal{Z}(u)$}{
        Compute likelihood $p_\theta(y^* \mid z)$ via a forward pass of the likelihood network\;
        Compute prior $p_\psi(z \mid H(u))$ via a forward pass of the prior network\;
    }

   Compute the optimal posterior and stop gradients: 
    \[
    q^*(z \mid H(u), y^*) =
    \mathrm{sg}\!\left[
    \frac{p_\theta(y^* \mid z)\, p_\psi(z \mid H(u))}
    {\sum_{z'} p_\theta(y^* \mid z')\, p_\psi(z' \mid H(u))}
    \right]
    \]

    \textbf{Update likelihood network} $\theta$ by maximizing:
    \[
    \mathbb{E}_{q^*(z \mid H(u), y^*)}
    \big[\log p_\theta(y^* \mid z)\big]
    \]

    \textbf{Update prior network} $\psi$ by minimizing:
    \[
    \mathrm{KL}\big(q^*(z \mid H(u), y^*) \,\|\, p_\psi(z \mid H(u))\big)
    \]
}

\end{algorithm}

\subsection{Flexible Inference via Learned Prior}
During inference, the target item $y^*$ is unknown, so the posterior $q(z|H(u),y^*)$ is inaccessible.
Instead, \proposed relies on the learned prior distribution $p_{\psi}(z|H(u))$, as defined by the marginal formulation in Eq.~\ref{eq:marginal}.
Given a user's interaction history, the model computes the probability for each latent view and integrates the predictions across all views accordingly.
This resulting marginal prediction is expressed as $p(y |H(u)) = \sum_z p_\theta(y|z) \, p_\psi(z|H(u))$.
This unified formulation enables \proposed to integrate complementary knowledge arising from multi-view combinations within a single framework. 

At inference time, we consider two strategies: (1) selecting the most probable view $z^* = \arg\max_z p_\psi(z|H(u))$ and using $p_\theta(y|z^*)$ for efficient prediction, and (2) aggregating predictions across all views following Eq.~\ref{eq:marginal} to effectively leverage complementary knowledge.
This flexibility allows \proposed to balance efficiency and accuracy depending on computational constraints.

\section{Experiments}
\label{sec:exp}
\subsection{Experimental Settings}
\label{subsec:expset}
\subsubsection{Datasets}
In our experiments, we utilize three datasets: \textbf{\amazonb}, \textbf{\amazons}, and \textbf{\yelp}.
The \amazon datasets \cite{he2016ups} are obtained from \textit{Amazon.com} for product recommendations, while the \yelp dataset comprises a collection of user ratings and reviews for business recommendations.
Dataset statistics are summarized in Table~\ref{tbl:datastats}.
For each dataset, we apply 5-core filtering and adopt a leave-one-out setting to split it into training, validation, and testing sets.
Specifically, for each user's interaction history, the second-to-last and last items are allocated to the validation and testing sets, respectively, while all preceding items are used for training.
We also limit each user's maximum sequence length to 20 interactions.
All settings are identical to those in previous studies~\cite{kang2018self,zhou2020s3,hua2023index,rajput2024recommender}.

\begin{table}[h]
    \centering
    \caption{Statistics of the user-item interaction datasets.}
    \label{tbl:datastats}
    \setlength{\textfloatsep}{1pt}
    \resizebox{\linewidth}{!}{
    \begin{tabular}{rrrr}
    \toprule
     & \textbf{Amazon Beauty} & \textbf{Amazon Sports} & \textbf{\yelp}\\
    \midrule
     {\#Users} & 22,363&25,598& 30,431\\
     {\#Items} &12,101&18,357&20,033\\
     {\#Interactions} &198,502&296,337&316,354\\
     {Avg\#Inter./User} & 8.9&8.3&10.4\\
     {Sparsity(\%)} &99.93&99.95&99.95\\
     \bottomrule
    \end{tabular}
    }
\end{table}

\begin{table*}[t]
\centering
\caption{Recommendation accuracy of all the baselines and our methods. The best and second-best results are highlighted in bold and underlined, respectively. The superscript $\dagger$ reports the results from their papers. $*$ indicates the statistical significance of $p<0.01$ from the paired t-test with the best baseline method.
– denotes that the official implementation code is not available.}
\resizebox{\linewidth}{!}{
\begin{tabular}{lXXXXXXXXXXXX}
\toprule
\multirow{2}{*}{\textbf{Method}} & \multicolumn{4}{c}{\textbf{\amazonb}} & \multicolumn{4}{c}{\textbf{\amazons}} & \multicolumn{4}{c}{\textbf{\yelp}} \\ 
 & \textbf{H@5} & \textbf{N@5} & \textbf{H@10} & \textbf{N@10} & \textbf{H@5} & \textbf{N@5} & \textbf{H@10} & \textbf{N@10} & \textbf{H@5} & \textbf{N@5} & \textbf{H@10} & \textbf{N@10} \\
\midrule
\lightgcn &0.0312&0.0192&0.0532&0.0262&0.0215&0.0137&0.0364&0.0185&0.0277&0.0175&0.0467&0.0236\\
\midrule
\grurec$^{\dagger}$ &0.0164&0.0099&0.0283&0.0137&0.0129&0.0086&0.0204&0.0110&0.0152&0.0099&0.0263&0.0134 \\
\bertrec$^{\dagger}$ &0.0203&0.0124&0.0347&0.0170&0.0115&0.0075&0.0191&0.0099&0.0051&0.0033&0.0090&0.0045 \\
\sasrec$^{\dagger}$ &0.0387&0.0249&0.0605&0.0318&0.0233&0.0154&0.0350&0.0192&0.0162&0.0100&0.0274&0.0136 \\
\sthreerec$^{\dagger}$ &0.0387&0.0244&0.0647&0.0327&0.0251&0.0161&0.0385&0.0204&0.0201&0.0123&0.0341&0.0168 \\
\midrule
\tiger$^{\dagger}$ &0.0454&0.0321&0.0648&0.0384&0.0264&0.0181&0.0400&0.0225&0.0212&0.0146&0.0367&0.0194  \\
\pfivecid$^{\dagger}$ &0.0489&0.0318&0.0680&0.0357&0.0313&0.0224&0.0431&0.0262&0.0261&0.0171&0.0428&0.0225 \\
\pfivesid$^{\dagger}$ &0.0433&0.0299&0.0652&0.0370&0.0274&0.0193&0.0406&0.0235&0.0202&0.0131&0.0324&0.0170 \\
\pfivecsid$^{\dagger}$ &0.0355&0.0248&0.0545&0.0310&0.0043&0.0031&0.0070&0.0039&0.0021&0.0016&0.0056&0.0029 \\
\pfiveciid$^{\dagger}$ &{0.0512}&{0.0356}&{0.0732}&{0.0427}
&{0.0321}&{0.0227}&{0.0456}&{0.0270}& {0.0287}&{0.0195}& {0.0468}&{0.0254} \\
LC-Rec $^{\dagger}$ & {0.0443}& {0.0311}&{0.0610}&{0.0331}&{0.0304}&{0.0196}&{0.0451}&{0.0246}&{0.0230}&{0.0158}&{0.0359}&{0.0199}\\
EAGER $^{\dagger}$ &\textbf{0.0618}&\textbf{0.0451}&\textul{0.0836}&\textbf{0.0525}&0.0281&0.0184&0.0441&0.0236&0.0265&0.0177&0.0453&0.0242 \\
EAGER-LLM $^{\dagger}$ & {0.0548}& {0.0369}&{0.0830}&{0.0459}& \textul{0.0373}& \textul{0.0251}& \textul{0.0569}& \textul{0.0315}&{-}&{-}&{-}&{-}\\
\midrule
\proposed($z^*$) & {0.0560}& {0.0382}&{0.0831}&{0.0469}&
{0.0358}&{0.0240}&{0.0533}&{0.0296}&
 \textul{0.0349}$^*$& \textul{0.0233}$^*$& \textul{0.0536}$^*$& \textul{0.0293}$^*$\\
 
\proposed 
&\textul{0.0608} &\textul{0.0414}&\textbf{0.0878}$^*$&\textul{0.0510}
&\textbf{0.0392}$^*$&\textbf{0.0281}$^*$&\textbf{0.0592}$^*$&\textbf{0.0329}$^*$
&\textbf{0.0370}$^*$&\textbf{0.0247}$^*$&\textbf{0.0591}$^*$&\textbf{0.0317}$^*$\\

\bottomrule

\end{tabular}
}
\label{tbl:mainresults}
\end{table*}

\subsubsection{Evaluation setup}
To assess the recommendation performance, we employ two popular top-$N$ ranking metrics: 
(1) hit ratio (\textbf{H@}$\bm{K}$) and (2) normalized discounted cumulative gain (\textbf{N@}$\bm{K}$). 
In our experiments, we set $K$ to 5 and 10. 
For rigorous evaluation, we report full-ranking results across the entire item set rather than using a sample-based evaluation.
For the generation of a ranked list from a language model, we employ beam search with a beam~size~of~20.

\subsubsection{Baselines}
We compare various recommendation methods, categorized into three groups. 
The first group includes the representative collaborative filtering method.
\begin{itemize}[leftmargin=1.2em, itemsep=2pt, topsep=2pt]
    \item \textbf{\lightgcn}~\cite{he2020lightgcn} utilizes lightweight GCN encoders, which linearly propagate the interaction of users and items.
\end{itemize}
The second group includes autoregressive methods.
\begin{itemize}[leftmargin=1.2em, itemsep=2pt, topsep=2pt]
    \item \textbf{\grurec}~\cite{hidasi2018recurrent} is a GRU-based sequence model that predicts the next item given the user's sequential interaction history. 
    \item \textbf{\bertrec}~\cite{sun2019bert4rec} employs the transformer architecture for sequential recommendation.
    It adopts a mask prediction task for BERT to effectively model item sequences.
    \item \textbf{\sasrec}~\cite{kang2018self} is a sequential recommender that employs a unidirectional network, adopting a self-attention mechanism to model long-term dependencies in item sequences. 
    \item \textbf{\sthreerec}~\cite{zhou2020s3} enhances data representation in recommender systems by leveraging self-supervised learning methods to capture correlations between items and attributes. 
\end{itemize}
The last group is the generative recommendation approach.
\begin{itemize}[leftmargin=1.2em, itemsep=2pt, topsep=2pt]
    \item \textbf{P5-IDs}~\cite{hua2023index} presents various indexing approaches designed for sequential recommendation. 
    The \pfive model is trained using each indexing approach:
    \begin{itemize}
        \item \textbf{CID} constructs the index by hierarchical clustering~\cite{von2007tutorial} on the co-occurrence matrix.
        \item \textbf{SemID} constructs the index by directly utilizing items' hierarchical category information.
        \item \textbf{IID} assigns independent unique identifiers for  items.
        \item \textbf{CID+IID} (or \textbf{+SemID}) generates the index via a combination of CID and IID (or SemID) by appending IID (or SemID) to CID's final sequence.
    \end{itemize}
    \item \textbf{\tiger}~\cite{rajput2024recommender} adopts the generative retrieval for sequential recommendation. It introduces a semantic ID constructed using \rqvae to uniquely identify items. 
    \item \textbf{\eager}~\cite{wang2024eager} employs a two-stream generation architecture that leverages separate decoders for behavioral and semantic tokens.
    \item \textbf{LC-Rec}~\cite{zheng2024adapting} proposes a LLM-based framework that integrates semantic IDs with metadata through auxiliary learning tasks.
    \item \textbf{\eager-LLM}~\cite{hong2025eager} extends the \eager to an LLM-based architecture that jointly models behavioral and semantic indices.
  
\end{itemize}
Note that the baselines of \pfiveciid and \pfivecsid simply combine two different item indices. 
In contrast, \proposed integrates predictions across views within a unified probabilistic model.

\subsubsection{Experimental settings}
For the \rqvae used for item index construction, the encoder and decoder each contain five linear layers with ReLU activation.
The input embedding and latent vector dimensions are 768 and 32, respectively.
The codebook size and sequence length are fixed to 256 and 3 for all datasets.
Code vectors are initialized using k-means clustering with 100 iterations on the instances of the first batch.
We train \rqvae for 10,000 epochs using the AdamW optimizer~\cite{loshchilov2017decoupled} with a learning rate of 1e-3 and a batch size of 4,096, without dropout or batch normalization.
We pre-train the \pfive sequential recommender model for 20 epochs using AdamW with a batch size of 32 and a peak learning rate of $1\times10^{-3}$ under a linear scheduler.
The prior network is implemented as a single linear layer on top of the \tfive encoder and is trained for 20 epochs with a learning rate of $1\times10^{-4}$ under a linear scheduler.
All methods are evaluated using three different random seeds, and we report the average performance.
Statistical significance is assessed using paired t-tests ($p<0.01$) against the best-performing baseline.

\subsection{Experimental Results}
\label{subsec:expresults}
\subsubsection{Effectiveness of \proposed framework}
\label{subsubsec:maineval}
Table~\ref{tbl:mainresults} presents the overall recommendation results across three datasets.
\proposed consistently achieves the highest performance in most metrics, demonstrating the effectiveness of variational integration across multiple views capturing complementary knowledge.
Early generative models such as \tiger and \pfive-IDs rely on a fixed index representation, which limits their ability to exploit complementary knowledge encoded in different views.
As discussed in Section~\ref{subsec:motivanal}, each view encodes distinct information sources: behavioral co-occurrence patterns from collaborative interactions and semantic attributes derived from item metadata.
Single-view models observe only one of these perspectives, resulting in incomplete user preference modeling.
A straightforward extension is to concatenate multiple indices (e.g., CID + SemID, CID + IID) into a unified sequence.
However, these approaches are insufficient to leverage complementary knowledge, as the consecutive generation of hierarchical index tokens disrupts the decoding process and weakens contextual integration across views.
To address these limitations, the dual-decoder framework \eager separates behavioral and semantic decoding streams and merges their outputs via log-probability-based reranking at inference.
While this design learns distinct views, it cannot identify their relative contribution under different user contexts.

In contrast, \proposed formulates this integration through a probabilistic framework, where a learned prior estimates the view-level contribution as a latent variable, allowing the model to capture user-specific weighting patterns.
This mechanism allows the model to adaptively determine which view is more appropriate for each user during training, enabling flexible and context-aware integration of complementary knowledge.
Even when using only a single inferred view (\proposed ($z^*$)), the model remains effective, indicating that the learned prior captures meaningful evidence of view importance.

Furthermore, despite using a much smaller backbone (\tfive-small), \proposed still outperforms larger LLM-based models built on Llama-7B~\cite{touvron2023llama}, such as LC-Rec and \eager-LLM, suggesting that effective multi-view probabilistic integration can be more influential than model scale in this setting.
Overall, these results highlight that \proposed unifies heterogeneous template-index combinations within a single probabilistic model, achieving adaptive integration of complementary knowledge across multiple views.

\begin{table*}[t]
\centering
\caption{
Ablation study on \textbf{index selection strategy} (top) and 
\textbf{prior aggregation with varying template numbers} (bottom). 
The upper block compares the performance when using a single index type (\ceid-only or \seid-only), 
while the lower block analyzes how different prior integration strategies and the number of templates 
($|\mathcal{T}|$) affect the final recommendation performance. 
}
\resizebox{1\linewidth}{!}{
\begin{tabular}{lXXXXXXXXXXXX}
\toprule
\multirow{2}{*}{\textbf{Setting}} & \multicolumn{4}{c}{\textbf{\amazonb}} & \multicolumn{4}{c}{\textbf{\amazons}} & \multicolumn{4}{c}{\textbf{\yelp}} \\ 
 & \textbf{H@5} & \textbf{N@5} & \textbf{H@10} & \textbf{N@10} & \textbf{H@5} & \textbf{N@5} & \textbf{H@10} & \textbf{N@10} & \textbf{H@5} & \textbf{N@5} & \textbf{H@10} & \textbf{N@10} \\
\midrule
\ceid-only                   & 0.0484 & 0.0324 & 0.0732 & 0.0405
                            &0.0321&0.0212&0.0482&0.0264
                            &0.0307&0.0204&0.0491&0.0263 \\
\seid-only                  &0.0505&0.0354&	0.0750&	0.0433
                            &0.0305&0.0204&0.0483&0.0261
                            &0.0227&0.0152&0.0368&0.0196 \\

\proposed($z^*$)            &\textbf{0.0560}&\textbf{0.0382}&\textbf{0.0831}&\textbf{0.0469}
&\textbf{0.0358}&\textbf{0.0240}&\textbf{0.0533}&\textbf{0.0296}
 &\textbf{0.0349}& \textbf{0.0233}& \textbf{0.0536}& \textbf{0.0293}\\
\midrule
Self-Consistency~\cite{wang2022self}&0.0565&0.0379&0.0832&0.0465
                            &0.0351&0.0232&0.0528&0.0289
                            &0.0342&0.0227&0.0538&0.0290 \\
Uniform prior               &0.0576&0.0388&0.0831&0.0469
                            &0.0363&0.0241&0.0538&0.0297
                            &0.0356&0.0235&0.0557&0.0299 \\
\proposed ($|\mathcal{T}| = 2$)&0.0599&0.0409&0.0857&0.0500
                            &0.0388&0.0277&0.0589&0.0325
                            &0.0364&0.0243&0.0578&0.0313 \\
\proposed ($|\mathcal{T}| = 4$)&0.0605&0.0412&0.0864&0.0504
                            &0.0389&0.0279&0.0590&0.0326
                            &0.0368&0.0246&0.0580&0.0314 \\

\proposed ($|\mathcal{T}| = 8$)&0.0606&0.0413&0.0874&0.0508
                            &0.0391&0.0280&0.0591&0.0328
                            &0.0370&0.0246&0.0588&0.0315 \\
\proposed ($|\mathcal{T}| = 10$)            &\textbf{0.0608}&\textbf{0.0414}&\textbf{0.0878}&\textbf{0.0510}
                            &\textbf{0.0392}&\textbf{0.0281}&\textbf{0.0592}&\textbf{0.0329}
                            &\textbf{0.0370}&\textbf{0.0247}&\textbf{0.0591}&\textbf{0.0317} \\
\bottomrule
\end{tabular}
}
\label{tbl:ablation}
\end{table*}

\subsubsection{Ablation study of multi-view integration}
\label{subsubsec:ablation}
To verify the effectiveness of each component, we perform a detailed ablation study on our proposed multi-view integration framework.
Table~\ref{tbl:ablation} shows the ablation results on index selection and view aggregation.

In the upper block, we examine the impact of using a single index type by comparing \ceid-only and \seid-only variants.
Both single-view variants exhibit clear performance degradation compared to \proposed($z^*$), indicating that each index type reflects different information sources when the index representations are generated.
While \ceid encodes behavioral co-occurrence patterns derived from user–item interactions, \seid abstracts semantic attributes obtained from textual and categorical metadata.
Although both indices are trained within the same sequential modeling framework and observe the same interaction sequence, the heterogeneity in their information sources leads each to capture distinct and partially overlapping aspects of user preference.
By contrast, integrating these heterogeneous views through the learned prior enables \proposed to infer which representation is more informative for a given user context, allowing complementary utilization of diverse views.

In the lower block, we further compare different strategies for aggregating multiple views.
Both the \textit{Self-Consistency} baseline~\cite{wang2022self}, which aggregates predictions via majority voting, and the \textit{Uniform prior}, which averages all views with equal weights, can combine prediction signals to some extent.
However, since neither approach explicitly models the probabilistic relationship among views, they implicitly assume that all inference trajectories are equally informative.
As a result, these methods lack the ability to distinguish which view provides more informative signals across users.

In contrast, \proposed assigns a user-conditioned probability distribution over predefined inference trajectories derived from template–index combinations.
Rather than functioning as a simple scoring or voting mechanism, the learned prior models the relative suitability of each trajectory given the user’s interaction history and uses this distribution to guide view integration during inference.
From this perspective, the proposed framework can be interpreted as a form of test-time scaling, where multiple inference trajectories are considered at inference time but are integrated through a learned probabilistic process rather than heuristic aggregation such as majority voting.
Moreover, this probabilistic modeling provides a structured signal over inference trajectories, suggesting potential extensions toward reinforcement learning settings.

Moreover, as the number of available prompt templates increases, performance consistently improves.
This trend indicates that \proposed effectively leverages diverse template-index combinations as complementary inference trajectories, and integrates them through the learned prior rather than over-specializing to any single view.

\subsubsection{Prior Distribution Analysis}
\begin{figure}[h]
    \centering 
    \includegraphics[width=\linewidth]{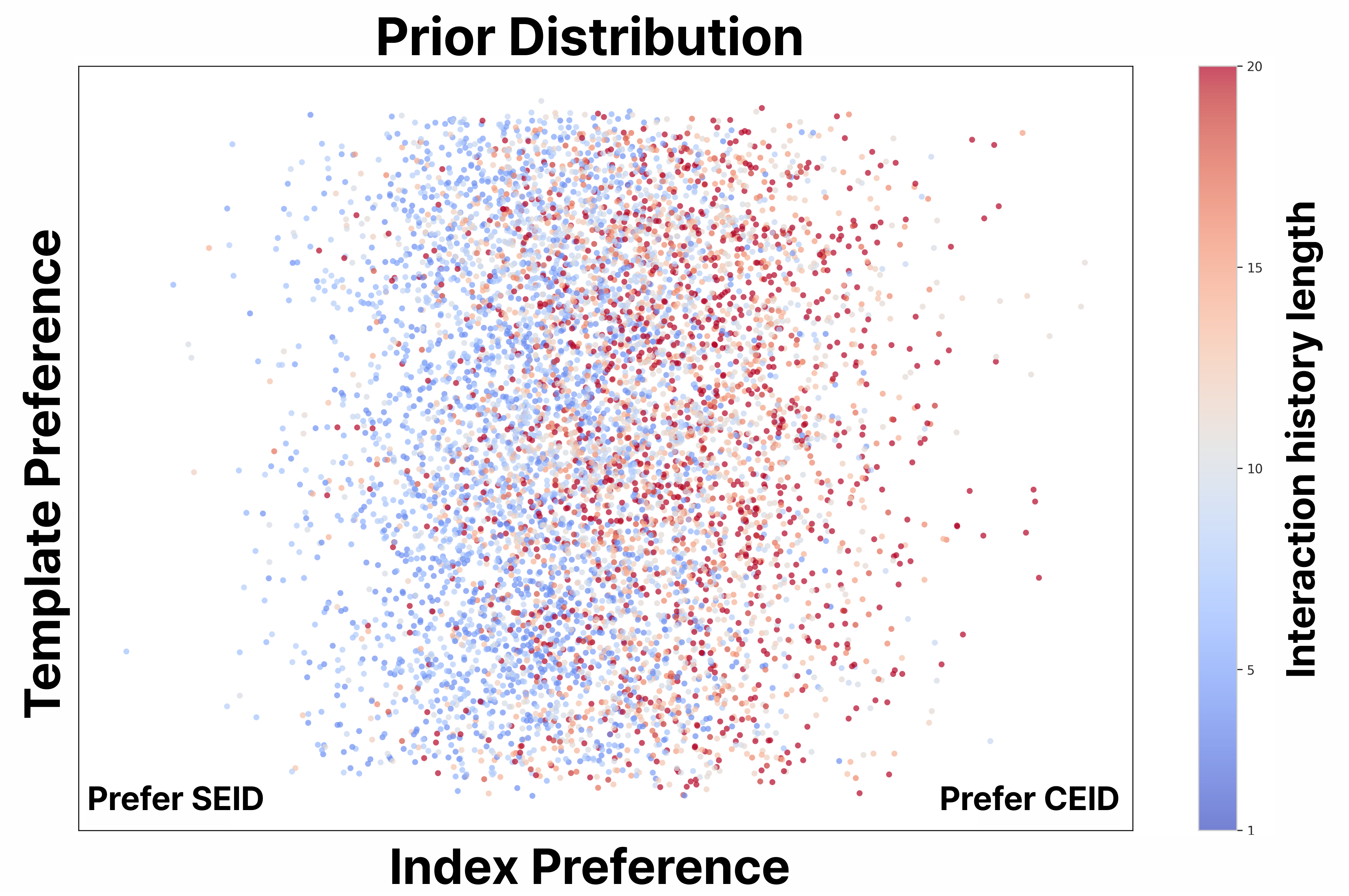} 
    \caption{Two-dimensional visualization of the learned prior distribution $p_{\psi}(z \mid H(u))$ across users on the Beauty dataset.}
    \label{fig:prior_analysis} 
\end{figure}

To analyze how the learned prior distribution $p_\psi(z \mid H(u))$ varies across users, we visualize user-specific priors in a two-dimensional space, as shown in Figure~\ref{fig:prior_analysis}.

The x-axis (Index Preference) reflects the relative preference over index types, indicating whether the learned prior distribution favors \seid- or \ceid-based views.
The y-axis (Template Preference) captures how the prior distribution is allocated across prompt templates, obtained by projecting the template-wise prior distribution into one dimension via t-SNE.
Each point corresponds to an individual user and is colored by the length of the interaction history, where blue denotes cold-start users (with fewer interactions) and red denotes users with richer interaction histories.

A clear pattern emerges along the index preference axis.
Users with richer interactions tend to exhibit priors biased toward \ceid-based views, whereas cold-start users are more likely to favor \seid-based views.
This observation suggests that the learned prior captures an implicit pattern revealed in user interaction data.
In particular, collaborative signals encoded in \ceid become more informative as the interaction history grows, while semantic signals encoded in \seid play a more prominent role for users with limited histories.

In contrast, no clear separation is observed along the template preference axis.
Users are broadly distributed across different prompt templates, indicating that prompt templates do not form a dominant axis for distinguishing user groups in the learned prior distribution.
Rather than functioning as fixed user-specific preferences driven by prompt phrasing (e.g., ``predict next item'' versus ``what to buy''), templates primarily serve to induce diverse inference trajectories.
These inference trajectories are subsequently weighted and integrated by the learned prior during inference, which can be viewed as a form of test-time scaling, as discussed in Section~\ref{subsubsec:ablation}.

Overall, this analysis shows that the learned prior captures user-specific preferences over index types derived from different knowledge sources, while leveraging prompt templates to support effective multi-view integration of complementary knowledge rather than over-specializing to particular prompt templates.

\subsubsection{Integration with alternative indexing methods}
\label{subsubsec:prior-dist}
To verify the compatibility and generality of our framework, we replace \ceid and \seid with CID and SemID from \pfive-IDs~\cite{hua2023index} and evaluate the performance in Table~\ref{tbl:idx-beauty}.
Despite differences in how these indices are generated, our framework consistently achieves comparable results.
This shows that the proposed variational integration mechanism is not tied to a specific indexing design, but rather focuses on modeling the relationships and complementary information across heterogeneous views.
The results highlight that \proposed can flexibly incorporate various indexing strategies and effectively benefit from recent advances in generative indexing techniques~\cite{chen2024enhancing, zhu2024cost, wang2024learnable, si2024generative, zheng2025universal}, confirming its broad applicability and extensibility.

\begin{table}[h]
\centering
\small
\caption{
Evaluation of alternative indexing method on \amazonb to assess the framework’s compatibility.
}
\resizebox{\linewidth}{!}{
\begin{tabular}{l|cccc} 
\toprule
\textbf{Indexing Method} & H@5 & N@5 & H@10 & N@10 \\
\midrule
CID & 0.0489 & 0.0318 & 0.0680 & 0.0357 \\
SemID & 0.0433 & 0.0299 & 0.0652 & 0.0370 \\
\proposed (CID+SemID) & \textbf{0.0587} & \textbf{0.0421} & \textbf{0.0818} & \textbf{0.0495} \\
\bottomrule
\end{tabular}
}
\label{tbl:idx-beauty}
\end{table}

\subsubsection{Inference time analysis}
\label{subsubsec:speed}
To analyze the computational efficiency and scalability of our framework, we measure the average inference time per user on a single RTX~4090 GPU while varying the number of prompt templates and item indices, as reported in Table~\ref{tab:speed}.
During inference, \proposed evaluates each of the $B$ (beam size) retrieved items across all combinations of item indices and prompt templates, resulting in a complexity of $O(B \times |\mathcal{I}| \times |\mathcal{T}|)$.
Relative to prior encoder-decoder generative recommenders, single-index baselines such as \tiger, and \pfive(IID, SemID, CID) correspond to the $|\mathcal{I}|=1, |\mathcal{T}|=1$ setting, whereas double-index baselines such as \pfiveciid, \pfivecsid and \eager correspond to the $|\mathcal{I}|=2, |\mathcal{T}|=1$ setting. 
Thus, compared with existing double-index baselines, the additional cost of \proposed mainly comes from using multiple prompt templates.
The prior inference step $p_\psi(z|H(u))$, which estimates the distribution over views, takes only about 0.01 seconds per user in our setup, indicating that prior estimation itself adds only a small overhead.
Table~\ref{tab:speed} shows that, as the number of templates increases, the inference time does not grow strictly linearly, indicating that multi-template inference can be implemented relatively efficiently through batch processing. 
This practical efficiency partially mitigates the additional cost introduced by broader view configurations. 
Consequently, \proposed supports flexible view configurations under different computational budgets, ranging from compact settings with lower inference cost to richer multi-view settings that can yield further performance improvements.

\begin{table}[h]
    \centering
    \small
     \caption{Inference time per user (in seconds), varying the number of prompt templates ($|\mathcal{T}|$) and item indices ($|\mathcal{I}|$).}
    \resizebox{\linewidth}{!}{
    \begin{tabular}{c|cccc|c}
    \toprule
    \textbf{$|\mathcal{I}|$} & \textbf{$\vert\mathcal{T}\vert$=1} & \textbf{$\vert\mathcal{T}\vert$=2}  &\textbf{$\vert\mathcal{T}\vert$=5}  &\textbf{$\vert\mathcal{T}\vert$=10} & \textbf{$p_{\psi}(z|H(u))$}\\
    \midrule
    1 & 0.05 & 0.08 & 0.13 & 0.26  & 0.01\\
    2 & 0.06 & 0.11 & 0.22 & 0.40  & 0.01\\
    \bottomrule
    \end{tabular}
    }
     \label{tab:speed}
\end{table}

\subsubsection{Analysis of indexing hyperparameters}
\label{subsubsec:hyperparam}
To examine the impact of codebook configuration, we conduct additional experiments on the Beauty dataset by varying the codebook size and code length, as summarized in Table~\ref{tab:code}.
Our results show that these variations have only minor effects on overall performance,
indicating that the proposed framework is stable under different index configurations.

\begin{table}[h]
    \centering
    \small
    \caption{Analysis of the impact of codebook size ($W$) and code length ($L$) on the Beauty dataset performance.}
    \label{tab:code}
    \resizebox{\linewidth}{!}{
    \begin{tabular}{c|ccccc}
    \toprule
    $(W, L)$ & \textbf{(128, 3)} & \textbf{(256, 3)} & \textbf{(512, 3)} & \textbf{(256, 2)} & \textbf{(256, 4)} \\
    \midrule
    H@5 & 0.059	& 0.061	& 0.057	& 0.059	& 0.059\\
    N@5 &0.040	& 0.041	& 0.038	& 0.038	& 0.039\\
    H@10 &0.084	& 0.087	& 0.083	& 0.084	& 0.085\\
    N@10 &0.049& 	0.051& 	0.047	& 0.047& 	0.049\\
    \bottomrule
    \end{tabular}
    }
\end{table}

\section{Conclusion}
\label{sec:conclusion}
This paper presented an analysis showing that complementary knowledge emerges when a model processes multi-view inputs from diverse template–index combinations.
Motivated by this finding, we proposed \proposed, a unified variational framework that integrates knowledge across multiple views within a single probabilistic model.
By learning a prior distribution over latent views conditioned on user history, the framework adaptively selects and aggregates complementary knowledge, improving recommendation consistency and overall performance.
Extensive experiments on real-world datasets verified the effectiveness of our framework.

\section*{Limitations}
Despite its effectiveness, this work has several limitations. 
Our experiments are centered on the T5-small backbone~\cite{ni2021sentence} for fair comparison with previous encoder-decoder generative recommenders, without a systematic scaling study over larger or decoder-only models. 
Still, as discussed in Section~\ref{subsubsec:maineval}, \proposed shows that effective integration of complementary knowledge can yield strong performance even against methods using substantially larger backbones. 
In addition, broader multi-view configurations increase inference cost, although Section~\ref{subsubsec:speed} shows that this overhead can be partially alleviated through efficient batching and flexible view selection. 
The textual diversity of views is also derived from a predefined set of prompt templates~\cite{geng2022recommendation}, rather than from explicitly optimizing template construction.
We adopt this design to ensure fair comparison with previous generative recommendation studies~\cite{hua2023index} under the same prompting setup, and show that even such simple template variations provide complementary signals that can be effectively integrated. 
Finally, the current study focuses only on collaborative and semantic index types, leaving broader information sources such as visual, price information, or external knowledge unexplored.
Since the framework is not inherently tied to these two index types, extending it to more diverse information sources remains an important direction for future work.
Despite these limitations, our results show that multi-view recommendation contexts can be explicitly defined and that the complementary knowledge they provide can be effectively integrated within a unified probabilistic framework.

\section*{Acknowledgments}
This work was supported by the IITP grants funded by the Korea government (MSIT) (No. RS-2020-II201361; RS-2024-00457882, AI Research Hub Project; IITP-2026-RS-2020-II201819).

\balance
\bibliographystyle{ACM-Reference-Format}
\bibliography{bibliography}

\end{document}